\documentclass[aps,showpacs,twocolumn]{revtex4}

\usepackage{amsmath}
\usepackage{amsfonts}
\usepackage{amssymb}
\usepackage{epsfig}
\usepackage{bm}

\begin{document}

\title{Critical Temperature for Bose-Einstein condensation in quartic
       potentials}

\author{S. Gautam }
\email{sandeep@prl.res.in}
\author{D. Angom}
\email{angom@prl.res.in}

\affiliation{
   Physical Research Laboratory \\
   Navarangpura, Ahmedabad - 380 009
}

\begin{abstract}
The quartic confining potential has emerged as a key ingredient to obtain
fast rotating vortices in BEC as well as observation of quantum phase
transitions in optical lattices. We calculate the critical temperature
$T_c$ of bosons at which normal to BEC transition occurs for the quartic
confining potential. Further more, we evaluate the effect of finite
particle number on $T_c$ and find that $\Delta T_c/T_c$ is
larger in quartic potential as compared to quadratic potential for number
of particles $ < 10^5$. Interestingly, the situation is reversed if
the number of particles is $\gtrsim10^5$. 
\end{abstract}


\maketitle


\section{Introduction}

Particles in anharmonic potentials is a well studied example of chaotic
system. A simple but good model of a system which exhibits parametric
transitions from regular to chaos and vice-versa is the quartic oscillator.
The parameters which control the transitions are the coupling strengths of
the cross terms. The quartic oscillator besides being a case study
of chaotic system, plays a very important role in the guise of Higg's field,
through which all the known fundamental particles acquire finite masses. The
same potential also appears in optical lattices, where quantum phase transition
from superfluid to Mott insulator has been observed \cite{marcus-02}.
Optical lattices are regular intensity patterns of light created with
counter propagating laser beams. The laser beams have a Gaussian profile
and create an overlapping anharmonic potential across the optical lattice.
Recent theoretical studies show that realization of the quantum phase transition
can be more robust with a quartic potential \cite{olivier-06}.
Besides the quantum phase transitions, introducing a quartic confining
potential stabilizes fast rotating vortices in Bose-Einstein condensates
\cite{fetter-01,ghosh-04,kim-05,ionut-05,fetter-05,bargi-06}. 
In a recent work \cite{kling-07}, the condensation temperature and 
thermodynamic properties of a rotating ideal Bose gas in an anharmonic trap 
has been studied. However, in the experimental realization of the quadratic 
plus quartic confining potential, the observation of the fast rotating 
vortices eludes an unambiguous detection \cite{vincent-04}.

In this paper we calculate the critical temperature $T_c$ at which bosons
confined by a quartic potential condense. BEC in such a potential
was studied for low-dimensional systems \cite{vanderlei-91}. This calculation
requires the density of states. In our calculations we use the semiclassical
expression of the density of states, which is valid at higher energies. It
is however not appropriate to study low lying states. In particular, the
energy of the ground state is essential to estimate $T_c$ for finite
number of particles. To estimate the correction to $T_c$ in the finite particle
case, we calculate the ground state energy analytically
\cite{mathews}. The calculation is based on a method which optimizes the
matrix elements of the quartic potential Hamiltonian in the harmonic
oscillator basis. 

Our calculations show that $T_c$ in quartic confining potential is higher
than quadratic potential. This is perhaps to be expected, since $T_c$
varies as $N^{1/\alpha}$ when the density of states is
proportional to $\epsilon^{\alpha -1}$. In case of 3D harmonic oscillator
potential $\alpha=3$ whereas it is 9/4 in the case of 3D quartic potential. 
Hence, in the quartic potential case $T_c$ varies as $N^{4/9}$  compared to
$N^{1/3}$ in the case of harmonic potential. From our calculations, it is
evident that the cross terms increase $T_c$ in 2D as well as 3D quartic
potential. We find that $T_c$ rises by factor of 1.2 and 1.1 in the 3D and
2D potentials respectively. However, the experimental realizations of the 
trapping potentials, which are created from laser beams, are more appropriately
described without the cross terms.


\section{$T_c$ for 3D quartic potential trap}
The general form of the quartic oscillator potential is
$\lambda(\bm{r}.\bm{r})^2 $. It is homogeneous and has cross terms in the
Cartesian coordinate representation which couple motions along different axes. 
Potentials of this form occur in optical lattices, where counter propagating 
lasers create undulating patterns of standing radiation field. In one 
dimension, a pair of counter propagating Gaussian laser beams along $z$-axis 
of intensity profile $I_0\exp(-2r^2/w)$ creates an array of periodic intensity 
minima and maxima. These are located along the $z$-axis. Depending on the 
detuning of the laser, the atoms are attracted to the intensity minima or 
maxima. Usually the wavelength of the laser $\lambda$ is much smaller than the 
beam width $w$. To a very good approximation, the potential across a surface 
normal to the laser beam
\begin{equation}
\label{eq.a.a}
 V(r) = I_0\left (\frac{-2r^2}{w} + \frac{4r^4}{w^2}\right ).
\end{equation}
Tuning the parameters of the laser beams, it possible to retain only the
quartic term. For simplicity, neglecting the cross terms, in three dimension
\begin{equation}
\label{eq.a.b}
 V(x, y, z)=\lambda(x^4+y^4+z^4).
\end{equation}
The eigen energies of the corresponding Hamiltonian is the sum of eigen
values corresponding to each dimension. For the one dimensional quartic
oscillator, eigen values can be calculated by minimizing the expectation 
of the Hamiltonian in the basis states of harmonic oscillator of appropriately 
chosen frequency\cite{mathews}. Generalizing the result to three 
dimensional case, the eigen energy
\begin{equation}
\label{eq.a.c}
   \epsilon(n_1,n_2,n_3)= 1.389\sum_{i=1}^3(n_i+\frac{1}{2})^\frac{4}{3}
                        \left(\frac{\lambda{\hbar}^4}{m^2}\right)^\frac{1}{3}.
\end{equation}
Since $1.389(\lambda\hbar^4/m^2)^{1/3}$ has the dimensions of energy
we can represent this factor by $\hbar\omega$, then
\begin{equation}
\label{eq.a.d}
  \epsilon(n_1,n_2,n_3)=\sum_{i=1}^3(n_i+\frac{1}{2})^\frac{4}{3}\hbar\omega .
\end{equation}
We now determine the number of states $G(\epsilon)$ with energy less than a
given value $\epsilon$. For energies large compared to $\hbar\omega$, we may 
treat $n_i$'s as continuous variables and neglect the ground state energy.
To calculate $G(\epsilon)$, we introduce a coordinate system in terms of the 
three variables $\epsilon_i=n_i^{4/3}\hbar\omega$. In this coordinate system  
$\epsilon=\epsilon_1+\epsilon_2+\epsilon_3$ defines a surface of constant energy
$\epsilon$. Then $G(\epsilon)$ is proportional to the volume in the first 
octant bounded by the surface
\begin{equation}
 \label{eq.a.e}
  G(\epsilon)=\frac{27}{64(\hbar \omega)^{9/4}}\int_0^{\epsilon}
              {\epsilon_1}^{-1/4}d{\epsilon_1} \int_0^{\epsilon- \epsilon_1}
              \!\!\!\!\!\!\!\!\!\!{\epsilon_2}^{-1/4}d\epsilon_2
              \int_0^{\epsilon-\epsilon_1-\epsilon_2}
              \!\!\!\!\!\!\!\!\!\!\!\!\!\!\!\!\!\!
              \epsilon_3^{-1/4}d{\epsilon_3}.
\end{equation}
To evaluate the integral we use the relation
$\int_0^u{x}^{\nu-1}( u-x) ^{\mu-1} =u^{\mu+\nu-1}B(\mu,\nu)$, where
$B(\mu,\nu)= \Gamma(\mu)\Gamma(\nu)/\Gamma(\mu+\nu)$ \cite{grad}. Then the
density of states
\begin{equation}
  \label{eq.a.f}
  g(\epsilon)=\frac{dG(\epsilon)}{d\epsilon}=0.6852\frac{\epsilon^{5/4}}
              {(\hbar\omega)^{9/4}}.
\end{equation}
This expression for density of states is used to calculate $T_c$.
For bosons, the total number of particles occupying the excited states is
\begin{equation}
  \label{eq.a.g}
  N_{\rm exc}=\int_0^{\infty}\frac{g(\epsilon)}{e^{(\epsilon-\mu)/kT}-1}
               d\epsilon.
\end{equation}
At critical temperature $\mu \rightarrow 0$ in the case of bosons and 
$N_{\rm exc}$ is equal to the total number of bosons $N$. Evaluating the 
integral gives $T_c$ in terms of the number of bosons
\begin{equation}
   \label{eq.a.h}
   kT_c=\frac{N^{4/9}\hbar\omega}{\left[0.6852\Gamma(9/4)
        \zeta(9/4)\right] ^{4/9}}.
\end{equation}
The corresponding expression for three dimensional isotropic harmonic
oscillator potential is \cite{pethick}
\begin{equation}
  \label{eq.c.c}
  kT_c=\frac{\hbar \omega_o N^{1/3}}{[\zeta(3)]^{1/3}}.
\end{equation}
From eq.~(\ref{eq.a.h}) and eq.~(\ref{eq.c.c}), the ratio of the critical
temperatures in the two potentials is
\begin{eqnarray}
  \label{eq.a.i}
   \frac{( T_c)_{\rm quartic}}{( T_c)_{\rm harmonic}} &= &
   \frac{N^{4/9}\zeta(3)^{1/3}}{N^{1/3}\left[ 0.6852\Gamma(9/4)\zeta(9/4)
   \right]^{4/9}} \nonumber \\
   &=&1.006N^{1/9}.
\end{eqnarray}
Where we have taken $\hbar\omega=\hbar\omega_o$ to obtain the ratio.
The ratio is proportional to $N^{1/9}$, which means that for $10^7$ atoms,
$T_c$ in the case of 3D quartic potential trap is approximately six times
higher than that of the 3D isotropic harmonic trap.
\begin{figure}
  \includegraphics[width=6cm,angle=-90]{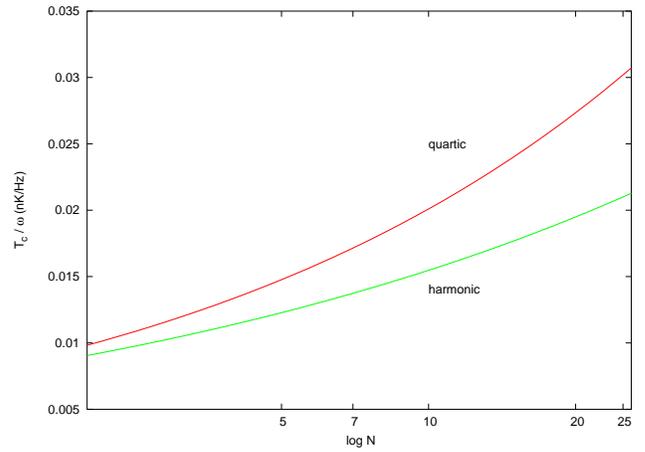}
  \caption{Ratio of the critical temperature ({$T_c$}) to angular frequency
           ({$\omega_o$}) as a function of the number of bosons on the log 
           scale. Here we have taken 
           $1.389(\lambda\hbar^4/m^2)^{1/3} =\hbar\omega_o$ ($\omega_o$ is
           angular frequency of isotropic harmonic oscillator). When 
           $\log N=7$ , the $T_c$ for the 3D quartic potential
           is approximately six times higher than the 3D isotropic 
           harmonic potential.}
  \label{fig.1}
\end{figure}


\section{Effect of finite particle number}

\begin{figure}
   \includegraphics[width=6cm,angle=-90]{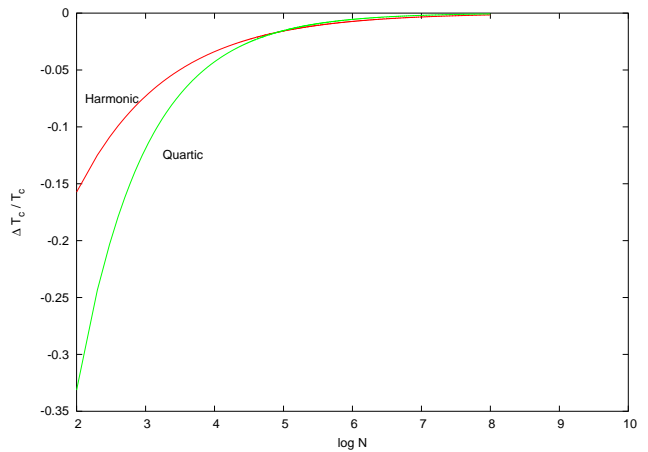}
   \caption{Fractional change in critical temperature ($\Delta T_c/T_c$) as
            a function the logarithm of the number of bosons ($\log N$ ). 
            The green and red colored plots correspond to the quartic and 
            quadratic potentials respectively. Below $\log (N)=5$,
            that is $N\thickapprox80,000$, the 3D quartic potential
            has larger fractional change. Above this point the 3D isotropic
            harmonic potential has marginally higher fractional change.
            }
\label{fig.2}
\end{figure}

The expression of $T_c$ in the previous section is with the approximation
that the ground state energy is zero, which is a valid approximation when the
system has large number of bosons. For finite number of bosons, the zero
point energy causes a change in the value of $T_c$.  For the 3D quartic 
potential, the ground state energy is \cite{mathews}
\begin{equation}
  \label{eq.a.j}
  \epsilon_{\rm min}=2.41\left( \frac{\lambda\hbar^4}{m^2}\right)^{1/3}.
\end{equation}
This should be equal to the change in the chemical potential at the critical 
temperature, that is $\Delta\mu=\epsilon_{\rm min}$. As the total number of 
bosons is fixed
\cite{pethick}
\begin{equation}
   \label{eq.a.k}
   dN=\left( \frac{\partial N}{\partial T}\right) _{\mu}dT+
\left( \frac{\partial N}{\partial{\mu}}\right)_T d\mu=0.
\end{equation}
This implies
\begin{equation}
   \label{eq.a.l}
   \left( \frac{\partial \mu}{\partial T}\right)_N= -\left( \frac{\partial N}
   {\partial T}\right) \left(\frac{\partial N}{\partial \mu} \right)^{-1}_T.
\end{equation}
Using the expression of $N$ at temperatures slightly above $T_c$
\begin{equation}
  \label{eq.a.m}
  N=C_{\alpha}\int_0^{\infty}\frac{\epsilon^{\alpha-1}}{e^{(\epsilon-\mu)/kT}
    -1}d\epsilon.
\end{equation}
This relation is obtained by substituting the general expression for the
density of states i.e. $g(\epsilon)=C_\alpha \epsilon^{\alpha-1}$ in
eq.~(\ref{eq.a.g}). Here $C_\alpha$ is a constant whose value depends on the
form of the trapping potential. Then from eq.~(\ref{eq.a.l}) and 
eq.~(\ref{eq.a.m}) we get
\begin{equation}
  \label{eq.a.n}
  \left( \frac{\partial \mu}{\partial T}\right)_N=-\alpha
  \frac{\zeta(\alpha)}{\zeta(\alpha-1)}k.
\end{equation}
In this expression $\alpha$ should be greater than 2, otherwise the
relation is not valid since $\zeta(1)$ diverges. Using this expression,
the change in the critical temperature due to the finite particle number is
\begin{equation}
   \label{eq.a.o}
   \Delta T_c=-\frac{\zeta(\alpha-1)}{\alpha\zeta(\alpha)k}\Delta\mu.
\end{equation}
In the case of 3D quartic potential $\alpha$ is equal to $9/4$. Then
\begin{eqnarray}
   \label{eq.a.p}
   \Delta T_c  & = & \frac{-4\zeta(5/4)}{9\zeta(9/4)k}\Delta\epsilon_{\rm min}
                \nonumber \\
               & = &\frac{-1.071\zeta(5/4)}{\zeta(9/4)k}
                    \left(\frac{\lambda\hbar^4}{m^2} \right)^{1/3}.
\end{eqnarray}
A relative measure of the effect of zero point energy on the critical
temperature is the fractional change of the critical temperature. It is
the ratio between $\Delta T_c$ and $T_c$, for the present case
\begin{equation}
   \label{eq.a.q}
   \frac{\Delta T_c}{T_c}=\frac{-0.9054\zeta(5/4)}{\zeta(9/4)}
        \left( \frac{\lambda\hbar^4}{m^2}\right)^{1/3}
        \frac{\left( \Gamma(9/4)\zeta(9/4)
        \right)^{4/9}}{\hbar\omega N^{4/9}}.
\end{equation}
Noting that $\left( \lambda\hbar^4/m^2\right)^{1/3}  $ is equivalent to
$\hbar\omega/1.389$ we get
\begin{equation}
\label{eq.a.r}
\frac{\Delta T_c}{T_c}=\frac{-0.6891\zeta(1.25)N^{-4/9}}
       {\zeta(2.25)^{5/9}}
\end{equation}
\begin{equation}
\label{eq.a.s}
        =-2.56N^{-4/9}.
\end{equation}
For the 3D isotropic harmonic potential, the fractional change of the 
critical temperature is
\begin{equation}
\label{eq.c.a}
  \frac{\Delta T_c}{T_c}=-0.73N^{-1/3}.
\end{equation}
If we compare eq.~(\ref{eq.a.s}) and eq.~(\ref{eq.c.a}), we find that the
percentage decrease in $T_c$ is larger in the case of 3D quartic potential 
trap for number of particles  $\lessapprox80,000$. But the scenario is
reversed for number of particles $>80,000$. This is also
evident from fig.~(\ref{fig.2}) where cross over point corresponds to
the number of particles $\thickapprox80,000$.


\section{Effect of cross terms}
\subsection{3D case}
Consider the general form of the quartic potential, as mentioned earlier
\begin{equation}
\label{eq.a.t}
  V(\bm{r})=\lambda(\bm{r}.\bm{r})^2.
\end{equation}
In optical traps, it is possible to create confining potentials which are
approximately close to this form but a truly spherically symmetric one is not
realizable. The difficulty is in producing the cross terms of the potential,
for example, terms like $x^2y^2$ in Cartesian coordinate representation. The
absence and presence of the cross terms in quadratic and quartic potentials
respectively introduce a key difference between the dynamics in the two
potentials. In absence of the cross terms like $xy$, in quadratic potential,
a perturbation to the dynamics of a particle along an axis remains
confined along that axis. In contrast, it propagates to other axes in the
case of quartic potential. For condensates in traps, an important parameter
which reflects the effects of these terms is the critical temperature.

  Semiclassically, total number of states available to the system can be
obtained by dividing the total phase space volume by $h^3$, the volume
of a single state 
\begin{eqnarray}
   \label{eq.a.u}
   G(\epsilon) & = &\frac{1}{h^3}\int{d\mathbf{x}}\int{d\mathbf{p}}, \\
   \label{eq.a.v}
    & = &\frac{16\pi^2}{h^3}\int_0^{r^*}r^2dr\int_0^{p^*}p^2dp.
\end{eqnarray}
In the above equation $r^*$ and $p^*$ are  radial coordinate and momentum
corresponding to the classical turning point respectively. Transforming the
variable of integration from $p$ to $\epsilon$ (using the relation
$p^2/{2m}=\epsilon-V(\bm{r})$) we get
\begin{equation}
 \label{eq.a.w}
  G(\epsilon)=\frac{16\pi^2m}{h^3}\int_0^{r^*}r^2dr
  \int_0^{\epsilon^*}\sqrt{2m(\epsilon-V(\bm{r}))}d\epsilon.
\end{equation}
Thus the density of states is
\begin{equation}
\label{eq.a.x} g(\epsilon)=\frac{16\pi^2m}{h^3}\int_0^{r^*}r^2
   \sqrt{2m\left( \epsilon-\lambda r^4\right) }dr.
\end{equation}
Substituting $r^4=x$ we can transform the integral into a form which can be
evaluated analytically \cite{grad}
\begin{eqnarray}
 \label{eq.a.y}
   g(\epsilon)&=&\frac{4\pi^2m^{3/2}\sqrt{2\lambda}}{h^3}
               \int_0^{\epsilon/\lambda}\left( \sqrt{\frac{\epsilon}
               {\lambda}-x}\right) x^{-1/4}dx \\
 \label{eq.a.z}
  &=&\frac{4\sqrt{2}\pi^2m^{3/2}\Gamma( 3/2)\Gamma( 3/4)\epsilon^{5/4}}
   {h^3\lambda^{3/4}\Gamma(9/4) }.
\end{eqnarray}
Using this expression for the density of states in eq.~(\ref{eq.a.g}) we get
\begin{equation}
\label{eq.b.a}
  kT_c=\frac{N^{4/9}}{\left[ \Gamma(9/4)\zeta(9/4)\right]^{4/9} }
        \left[ \frac{h^3\lambda^{3/4}\Gamma(9/4)}
        {\Gamma(3/2)\Gamma(3/4)4\sqrt{2}\pi^2m^{3/2}}\right]^{4/9}.
\end{equation}
Comparing with eq.~(\ref{eq.a.h}) we can obtain the ratio of $T_c$ in the
two cases, with and without the cross terms, for the 3D isotropic quartic 
potential. It is found that $T_c$ with the cross terms is 1.2 times higher.
This rise in $T_c$  can be attributed to the contribution from the cross 
terms which were neglected while deriving eq.~(\ref{eq.a.h}). The
reason for the difference is, when $g(\epsilon)= C_\alpha \epsilon^{\alpha-1}$
then $T_c$ varies as $1/{C_\alpha}^{1/\alpha}$. Hence, the lower $T_c$ in 
3D quartic potentials without the cross terms is due to the higher value of 
$C_\alpha$. 

\subsection{2D case}
In the 2D case, neglecting the cross terms, the potential is of the form
\begin{equation}
 \label{eq.b.b}
  V(x,y)=\lambda(x^4+y^4).
\end{equation}
Using the same approach as adopted in the 3D case, the total number of states
available to the system is
\begin{eqnarray}
\label{eq.b.c}
 G(\epsilon)&=&\frac{9}{16(\hbar\omega)^{3/2}}
               \int_0^\epsilon{\epsilon_1}^{-1/4}d \epsilon_1
               \int_0^{\epsilon-\epsilon_1}\epsilon_2^{-1/4} d\epsilon_2 \\
  \label{eq.b.d}
 &=&\frac{3\Gamma(3/4)\Gamma(7/4)}{4(\hbar\omega)^{3/2}\Gamma(5/2)}
    \epsilon^{3/2}.
\end{eqnarray}
Thus the density of states 
\begin{equation}
\label{eq.b.e}
   g(\epsilon)=\frac{0.9531\epsilon^{1/2}}{(\hbar\omega)^{3/2}}.
\end{equation}
Substituting this expression of $g(\epsilon)$ in eq.~(\ref{eq.a.g}) we get
\begin{equation}
 \label{eq.b.f}
  kT_c=\frac{{\hbar\omega}N^{2/3}}{\left[ \zeta(3/2)\Gamma(3/2)
       0.9531\right]^{2/3}} .
\end{equation}
The corresponding expression when cross terms are considered is
\cite{vanderlei-91}
\begin{equation}
\label{eq.b.g}
  kT_c=\left[ \frac{Nh^2\sqrt {\lambda}}{2\pi^2m\Gamma(3/2)\zeta(3/2)}
       \right]^{2/3}.
\end{equation}
Comparing eq.~(\ref{eq.b.f}) and eq.~(\ref{eq.b.g}) we find that
$T_c$ with the cross terms in eq.~(\ref{eq.b.g}) is approx. 1.12 times 
higher than $T_c$ without the cross terms in eq.~(\ref{eq.b.f}).  Thus the
cross terms increase $T_c$ in 2D as well as 3D case.


\section{Conclusions}
Our calculations show that $T_c$ in the case of the 3D quartic
potential trap is higher than that of the isotropic harmonic potential
 trap. This is due to the form of the density of states $g(\epsilon)$,
which varies as $\epsilon^{5/4}$ and $\epsilon^2$ in 3D
isotropic quartic and quadratic trapping potentials respectively.
This implies lower density of states in quartic oscillator potential
compared to isotropic harmonic oscillator potential. However, more interesting
is the effect of the cross terms. In the 3D isotropic harmonic potential trap
the cross terms are absent, which is not the case for the 3D quartic 
potential trap. The cross terms tend to decrease the density of states and 
raise $T_c$. These terms increase $T_c$ by factor of 1.2 and 1.1 in the
3D and 2D quartic trap potentials respectively as compared
to the case without the cross terms. Experimentally, in
optical traps, the potentials without the cross terms are more
appropriate. We find  that the effect of finite particle number
is more pronounced in the 3D quartic potential when the number of particles 
is $< 10^5$. The situation is reversed when the number of 
particles is $\gtrsim10^5$. The cause of the reversal lies in the form
of the fractional change $\Delta T_c/T_c$ for the two potentials. The 
ratio of the fractional change between 3D isotropic quartic potential to 
harmonic potential is $3.51/N^{1/9}$. It is $\thickapprox 1$ 
for $N\thickapprox 10^5$, $>1$ for $N<10^5$ and $<1$ for $ N\gtrsim10^5$. 
Thus, when  $N < 10^5$ the constant factor is dominant in 
eq.~(\ref{eq.a.s}) and is responsible for the larger value of $\Delta T_c/T_c$ 
in quartic potential. But when $N\gtrsim 10^5$, $N^{-4/9}$ dominates and
$\Delta T_c/T_c$ of the quartic potential is lower than that of the
harmonic potential.


\end{document}